# Cu-TBPP and PTCDA molecules on insulating surfaces studied by ultra-high-vacuum non-contact AFM

L Nony[1,2], R Bennewitz[1], O Pfeiffer[1], E Gnecco[1], A Baratoff[1],
E Meyer[1], T Eguchi[3], A Gourdon[4] and C Joachim[4]

[1] Institute of Physics, University of Basel, CH-4056 Basel, Switzerland
[2] IBM Research Division, Zurich Research Laboratory, CH-8803 Rüschlikon, Switzerland
[3] Institute for Solid State Physics, The University of Tokyo, 5-1-5 Kashiwa-no-ha, Kashiwa, Chiba 277-8581, Japan
[4] CEMES-CNRS, 29 rue J Marvig, PO Box 4347, F-31055 Toulouse Cedex, France

**Abstract**
The adsorption of two kinds of porphyrin (Cu-TBPP) and perylene (PTCDA) derived organic molecules deposited on KBr and $Al_2O_3$ surfaces has been studied by non-contact force microscopy in ultra-high vacuum, our goal being the assembly of ordered molecular arrangements on insulating surfaces at room temperature. On a Cu(100) surface, well ordered islands of Cu-TBPP molecules were successfully imaged. On KBr and $Al_2O_3$ surfaces, it was found that the same molecules aggregate in small clusters at step edges, rather than forming ordered monolayers. First measurements with PTCDA on KBr show that nanometre-scale rectangular pits in the surface can act as traps to confine small molecular assemblies.

## 1. Introduction

Well ordered films of molecules on flat substrates have attracted much interest for the development of new electronic devices. Self-organization in growth processes and the possibility of controlling electronic properties raise the hope that functional electronic structures with molecular dimensions can be created. Model systems of molecules designed for these purposes have been studied by methods of surface science such as scanning tunnelling microscopy (STM) or low-energy electron diffraction on atomically smooth surfaces. In STM experiments, these molecules have been shown to exhibit fascinating properties such as electron tunnelling along a molecular wire [1], reconstruction of the underlying metal substrate [2], or spontaneous assembly into molecular wires of co-adsorbed molecules [3].

The majority of such studies have been performed using metallic substrates, the high electric conductivity of which could prevent their use for transistor-like devices. In order to separate the functional electronic systems from the substrate, molecules with spacer legs have been designed [1] and adsorption on thin insulating spacer films has been explored [4]. As more complex systems of structured molecular monolayers and metallic wiring on insulating substrates are developed, the development of scanning force microscopy (SFM) into a tool with similar resolution on molecular films to STM is highly desirable. Progress in dynamic force microscopy, in particular non-destructive imaging, usually referred to as non-contact AFM (nc-AFM), led to several studies where molecular resolution could be obtained based on the tip–sample interaction for distance control (see [5] and references therein). Beyond imaging, force microscopy experiments can also determine forces involved in the manipulation of molecules [6]. So far, most of the systems studied with molecular resolution have been performed on conducting substrates and compared to existing STM results. We report nc-AFM experimental studies of molecular arrangements on insulating substrates at room temperature. This investigation is motivated by a search for molecules which assemble and order on electrically insulating substrates owing to diffusion and intermolecular interaction rather than to chemical bonding to the crystalline substrate.

## 2. Experimental set-up

The dynamic force microscope used in this study is a home-built instrument described in [7]. We use silicon cantilevers with typical resonance frequencies and spring constants of $f_0 = 160$ kHz and $k = 24$ N m$^{-1}$ or $f_0 = 330$ kHz and $k = 40$ N m$^{-1}$, respectively. The tip is oscillating at a constant amplitude $A$ of typically 10–20 nm. All measurements are performed in ultra-high vacuum (UHV) with a base pressure $< 10^{-10}$ mbar and at room temperature. The topography of samples is usually recorded by scanning the surface using the interaction-shifted resonance frequency, $\Delta f$, for distance control. Simultaneously, the additional damping of the cantilever oscillation is recorded [8]. When working with a conducting sample, a bias voltage can be applied between the tip and the surface. Then the tunnelling current averaged over the tip oscillation can be used for distance control.

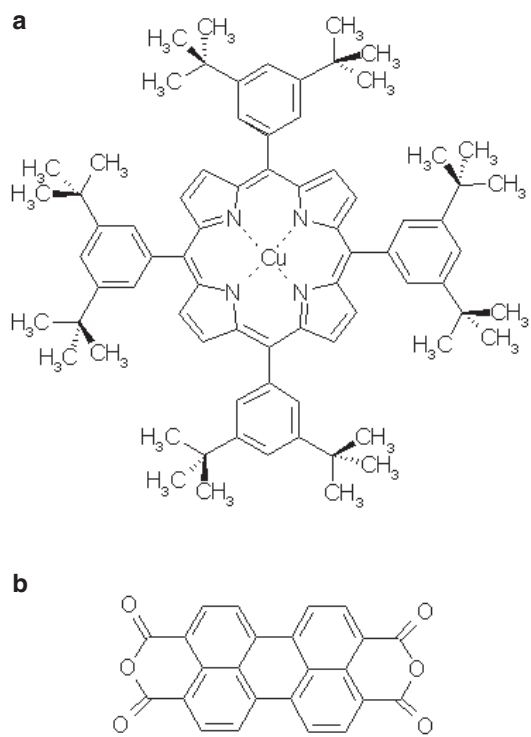

**Figure 1.** (a) Cu-tetra[3,5-di-tert-butylphenyl]porphyrin (Cu-TBPP), (b) 3,4,9,10-perylene tetracarboxylic dianhydride (PTCDA).

Two types of molecule have been investigated, namely Cu-tetra[3,5-di-tert-butylphenyl]porphyrin (Cu-TBPP, figure 1(a)) and 3,4,9,10-perylene tetracarboxylic dianhydride (PTCDA, figure 1(b)). Molecules are evaporated in UHV from a small Al$_2$O$_3$ crucible which is heated by a tantalum wire coil. The evaporator can be introduced into the vacuum chamber through a load-lock on top of a standard sample holder (Omicron Nanotechnologies). The deposition rates are calibrated using a Balzers quartz microbalance.

## 3. Experimental results

### 3.1. Cu-TBPP molecules on Cu(100)

We first deposited half a monolayer of Cu-TBPP molecules onto a Cu(100) surface at a substrate temperature of 150 °C. This preliminary study was motivated by the well ordered and stable monolayer islands observed for that system [9]. Figures 2(a) and (b) show nc-AFM images obtained using $\Delta f$ for distance control. Rectangular islands of molecules can be clearly identified in figure 2(a). The smaller image in figure 2(b) reveals rows of single molecules, with some blurry internal structure. For comparison, figure 2(c) shows an image obtained with the same oscillating tip using instead the time averaged tunnelling current for distance control. Keeping in mind the typical appearance with four protrusions per molecule on Cu(100) in figure 2(c) and in conventional STM images [9], one can even recognize similar intramolecular details in some molecules in the left-hand part of figure 2(b). The significant difference in contrast quality between STM and nc-AFM images of molecular films is found to be a reproducible phenomenon, as discussed in the last section.

### 3.2. Cu-TBPP molecules on a KBr surface with regularly spaced steps

In an attempt to obtain similarly ordered islands on an insulating surface, we deposited one-half of a monolayer of Cu-TBPP molecules onto a KBr surface pre-cleaved in UHV displaying a regular pattern of monatomic steps. These steps are part of a spiral which develops around a crystal dislocation when heating the sample in vacuum to 380 °C for 20 min [10, 11]. During subsequent deposition of the Cu-TBPP molecules, the surface was kept at room temperature.

Images of the surface are presented in figure 3 before (a) and after (b), (c) deposition of the molecules. The molecules decorate the steps by agglomerating into rounded interconnected aggregates. The height profile in figure 3(d) reveals that the height of the aggregates corresponds to a thickness of at least three monolayers. Looking at the images and at the height profile, it is difficult to judge whether there is any specific order in the molecular aggregates. Further attempts to obtain better resolution by choosing a more negative frequency shift, thereby decreasing the tip–sample distance, resulted in instabilities while scanning.

### 3.3. Cu-TBPP molecules on a Al$_2$O$_3$ surface with regularly spaced steps

A corresponding experiment has been carried out on a similarly structured Al$_2$O$_3$ surface, which was prepared by heating a sapphire crystal to 1400 °C in air for 24 h, transferring it to the vacuum chamber and flashing it three times to 1200 °C during 20 s. The surface then exhibits (0001) terraces separated by parallel steps, seen in figure 4(a), of height 0.7 nm, corresponding to one half of the lattice spacing along the $c$-axis. Some blurry features with sub-nm corrugation are also reproducibly imaged on the terraces.

Again, sub-monolayer coverage by Cu-TBPP molecules leads to aggregation into round clusters at the steps, as seen in figure 4(b). Although the chosen imaging parameters fail to provide any molecular resolution, some clusters are laterally shifted or picked up by the tip. Scanning the same frame once up and then down, six out of ten clusters have been removed, as seen in figure 4(c).

### 3.4. PTCDA molecules on a nano-structured KBr(001) surface

In an attempt to exploit the attraction of molecules toward steps for the directed growth of ordered molecular arrangements, we have used nanometre-scale rectangular pits of one atomic layer depth in a KBr(001) surface as template. Such pits, bounded by atomically straight steps [12], can be produced by careful electron irradiation. In order to increase the molecule–sample interaction, we chose PTCDA molecules [13] without spacer legs, whose aromatic cores should be closer to the surface compared to Cu-TBPP molecules.

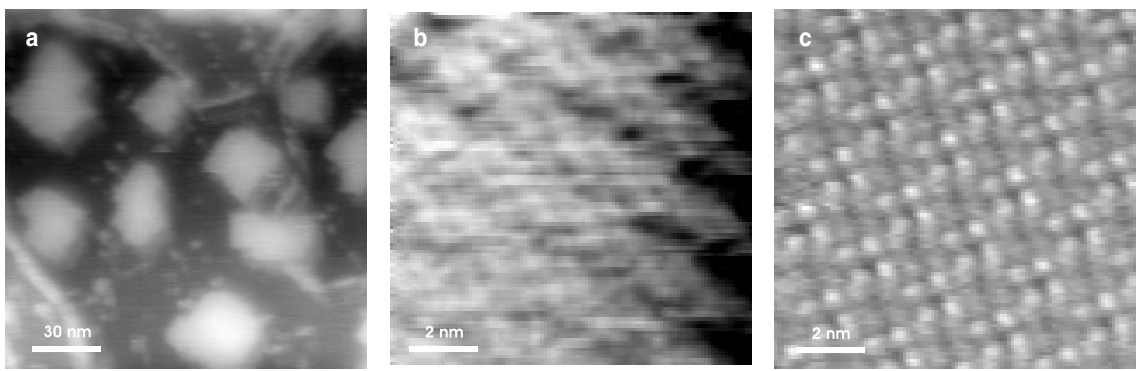

**Figure 2.** Topographic images of the Cu(100) surface partially covered by ordered monolayer islands of Cu-TBPP molecules. (a) Overview image obtained using the frequency shift $\Delta f$ for distance control (frame edge 127 nm, $f_0 = 161\,040$ Hz, $A = 9$ nm, $\Delta f = -50$ Hz). (b) Zoom showing rows of molecules and some internal structure (frame edge 8.1 nm, $f_0 = 161\,040$ Hz, $A = 9$ nm, $\Delta f = -50$ Hz). (c) High-resolution image recorded using the tunnelling current ($U_{\text{bias}} = 1$ V) for distance control (frame edge 8.1 nm). Four lobes can be recognized on each molecule.

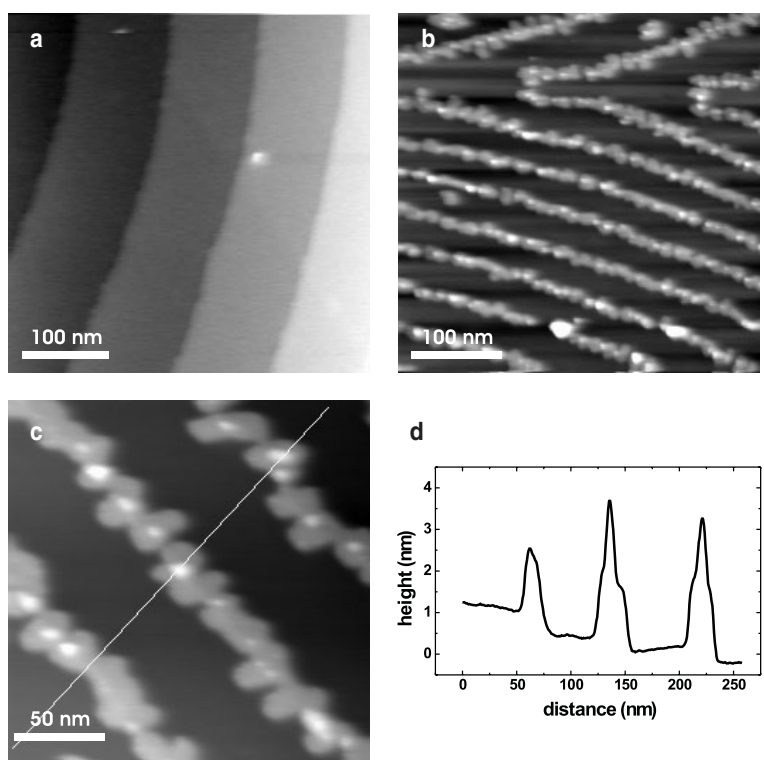

**Figure 3.** (a) Image of a KBr(001) surface (frame edge 418 nm) displaying terraces separated by regularly spaced monatomic steps which form part of a growth spiral. (b) The same surface after deposition of a submonolayer of Cu-TBPP molecules (frame edge 400 nm); the molecules decorate the step edges. (c) Detail of the decorated steps (frame edge 200 nm). The white line indicates the position of the height profile shown in (d). These nc-AFM images were recorded with identical parameters ($f_0 = 324\,331$ Hz, $A = 8.0$ nm, $\Delta f = -5$ Hz).

After deposition of one-tenth of a monolayer onto the nano-structured KBr surface, the molecules were found trapped in some of the pits. None were found on the KBr terraces. In the damping image, figure 5(b), seven such rectangular pits can clearly be recognized, five smaller ones through an enhanced damping signal across the pit, two larger ones through an enhanced damping signal along the step edges. The corresponding topographic image in figure 5(a) shows the two larger pits as depressions, while the smaller ones exhibit little contrast except close to some edges. The smaller pits seem to be filled with molecules, resulting in a nearly flat topographic appearance, but an enhanced damping. For the larger pits, only the edges seem to be decorated by molecules. This observation must be interpreted with care because a similar edge enhancement has been observed on KBr without any deposited molecules [14]. Thus it is not possible to unambiguously attribute the damping enhancement along the pit edges to molecules. However, enhanced damping in the inner part of a pit has not been previously reported and is probably due to molecules. The profiles in figures 5(c) and (d) quantify these observations. The measured depth of the larger pit is about 0.2 nm, less than the height of one monolayer of KBr. This reduced depth can be explained by convolution with the tip radius, which is of the same order as the lateral extension of the pit. The filled pit appears as a protrusion of about 0.05 nm, whereas the damping of the cantilever in the same region exceeds the intrinsic damping on the terraces by 0.25 eV per oscillation cycle. Attempts to obtain molecular resolution within the pits failed.

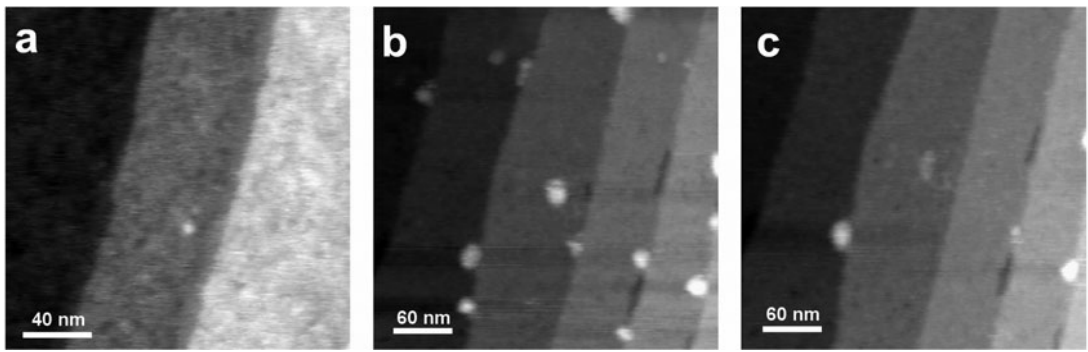

**Figure 4.** (a) As prepared $Al_2O_3(0001)$ surface (frame edge 200 nm, $f_0 = 324\,331$ Hz, $A = 8.0$ nm, $\Delta f = -7$ Hz); the step height is 0.7 nm. (b) and (c) Consecutively recorded images of the same surface after deposition of a sub-monolayer of Cu-TBPP molecules (frame edge 350 nm, $f_0 = 324\,331$ Hz, $A = 10.0$ nm, $\Delta f = -18$ Hz).

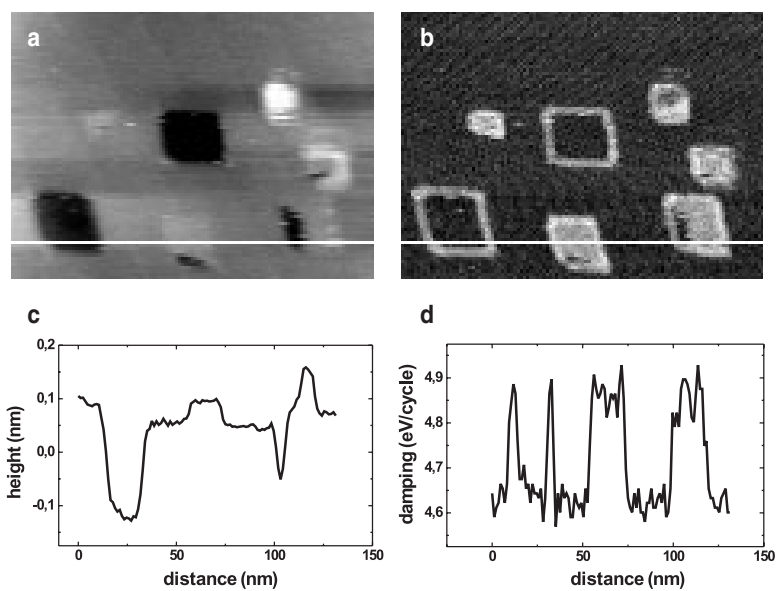

**Figure 5.** KBr(001) surface with electron irradiation-induced rectangular pits, covered with one-tenth of a monolayer of PTCDA molecules. (a) Topography signal recorded by nc-AFM (frame edge 132 nm × 98 nm, $f_0 = 162\,134$ Hz, $A = 17$ nm, $\Delta f = -15$ Hz). (b) Simultaneously recorded damping signal. The white bar indicates the location of the height profile in (c) and of the damping profile in (d). The damping has been quantified according to the procedure described in [14].

## 4. Discussion

The significantly lower quality of high-resolution nc-AFM images compared to STM images is a general finding in studies of molecular films on conducting substrates. In the following, we discuss three reasons for that difference, which all stem from the different interaction responsible for the contrast in the two methods. First, the short decay length of the tunnelling current selects very few atoms which probe the electronic overlap between tip apex and surface. In force microscopy, tip asperities as far as two or three nanometres from the surface can contribute to local force variations, thereby producing a more blurred contrast in high-resolution images. Tip sharpness is therefore a prerequisite to expect achieving a high vertical, but also lateral, contrast on heterogeneous surfaces. Second, unstable atomic configurations at the tip apex perturb nc-AFM experiments more than STM experiments. Such instabilities produce large fluctuations in the tunnelling current, but can eventually lead to more stable rearrangements; it is often a question of patient scanning until a stable tunnelling tip is formed, in a kind of self-stabilizing process. In contrast, atomic scale instabilities near the tip apex of an nc-AFM can produce variations of the total force, thus disturbing the imaging. One standard STM recipe for tip improvement, i.e. the application of voltage pulses, turns out to be difficult owing to interference with the maintenance of the tip oscillation, and is not feasible on insulating surfaces anyway. Another difficulty in nc-AFM on heterogenous surfaces is the variation of long-range forces between substrate or molecular islands on the one hand, and the tip on the other hand, for example due to work function variations [5, 15]. Feedback parameters suitable for high-resolution imaging on the substrate may be inadequate for the molecules. To overcome this problem, refined distance controllers have to be developed which selectively react to variations of long-range contributions to the total force.

Besides such general imaging issues in nc-AFM, a particular aspect highlighted by our work is the different behaviour of large aromatic molecules adsorbed on ionic insulators, as opposed to metals. The growth of ordered multilayers of such molecules on various substrates has mainly been documented by diffraction and other surface-averaging techniques [16]. In most cases, only quasi-epitaxy is observed, i.e. the molecular and substrate lattices are not commensurate, at least in one direction. In some cases, crystallites with the bulk stacking grow with molecular planes perpendicular to the substrate. These observations have been rationalized in terms of van der Waals, residual electrostatic and steric interactions between the molecules. Molecules with aromatic cores parallel to the substrate form ordered monolayers on noble metals at room temperature only if they are sufficiently large [17] or if they can form hydrogen bonds [18], thus ensuring strong enough intermolecular interactions. In addition, localized $\pi$-bonds might form if the aromatic core is close to a flat metal surface, as for PTCDA on Ag(111) [19]. Alternatively, a sufficient number of flexible 'legs' can adjust to more corrugated metal surfaces [9], although each leg is bonded by van der Waals interactions. In concert with intermolecular interactions, which affect the relative orientations of neighbouring molecules, binding to specific substrate sites leads to the commensurate monolayers characterized in detail by STM.

Such effects are expected to be weaker on ionic insulators. Their large energy gaps imply a weaker polarizability and weaker van der Waals interactions with a given molecule, roughly a factor of two smaller than for metals [20]. Bonds to specific sites might form with legs terminated by charged or polarizable groups, but this is not the case for Cu-TBPP, whereas the lone pairs on the oxygens at the ends of PTCDA seem insufficient. More suitable molecules in that respect are being investigated. In any case, the effective substrate corrugation experienced by isolated aromatic molecules without suitable legs is typically so weak that they easily diffuse at room temperature even on metals, at least along 'easy directions'. The same is expected even more on insulators. Intermolecular interactions, which can be stronger in the absence of metallic screening, are expected to favour the formation of ordered clusters. However, they might be difficult to detect by scanning probe microscopy because they would probably undergo appreciable lateral thermal motion and/or be easily dragged by the scanning tip. Indeed, it is difficult to account for the known number of deposited molecules on the basis of our nc-AFM images.

The observed aggregation into clusters at steps is probably due to the interaction of the lateral electric field generated at a step with the dipole induced in the extended $\pi$-system of each aromatic core. As a first approximation, we model a monatomic step on a metal surface as a continuous line of dipoles and on an ionic surface as a line of alternating charges with period $a$. The resulting electric field at a distance $x$ from the step exceeding the lattice constant varies as $\approx e/x^2$ in the former case, but as $\approx e/a^2 \exp(-2\pi x/a) \cos(2\pi y/a)$ in the latter one. Unless a multiple of the period along the step matches the intermolecular spacing preferred by intermolecular interactions, the second situation is less conducive to the formation of ordered arrangements along steps.

In addition to suggesting further experiments with molecules having side groups or legs better suited for site-specific adsorption on ionic substrates, our observations indicate that nano-structuring provides another promising avenue for binding compact islands to such substrates, as demonstrated for the PTCDA molecules trapped in pits created in the KBr(001) surface. The enhanced electrostatic interaction at steps probably acts as a barrier confining the molecules within the pits. Furthermore, the stronger field near corners of the pits should attract molecules more efficiently. In the preliminary results presented here, only small islands of about $20 \times 20$ nm$^2$ seem to be efficient traps. Based on the size of the unit cell of PTCDA as observed by STM [13], the number of molecules in one pit could be of the order of 300. Although the density and ordering of molecules within the pit is totally unknown at this time, the enhanced damping of the cantilever oscillation by 0.25 eV/cycle above hundreds of molecules in the filled pits is understandable considering the multitude of soft degrees of freedom of motion in this weakly bound molecular assembly. Nevertheless, the quantification of the damping signal as well as the underlying mechanisms is still under controversial discussion [21].

## Acknowledgments

This work was supported by the Swiss National Science Foundation, the Kommission zur Förderung von Technologie und Innovation, the National Center of Competence in Research on Nanoscale Science, and the European Union within the IST-FET 'Bottom-up nanomachine (BUN)' project.

## References

[1] Langlais V, Schlittler R, Tang H, Gourdon A, Joachim C and Gimzewski J 1999 *Phys. Rev. Lett.* **83** 2809
[2] Rosei F, Schunack M, Jiang P, Gourdon A, Laegsgard E, Stensgaard I, Joachim C and Besenbacher F 2002 *Science* **296** 328
[3] de Wild M, Berner S, Suzuki H, Yanagi H, Schlettwein D, Ivan S, Baratoff A, Güntherodt H-J and Jung T 2002 *Chem. Phys. Chem.* **3** 825
[4] Rauscher H, Jung T, Lin J-L, Kirakosian A, Himpsel F, Rohr U and Müllen K 1999 *Chem. Phys. Lett.* **303** 363
[5] Yamada H 2002 *Noncontact Atomic Force Microscopy* (*NanoScience and Technology*) ed S Morita, R Wiesendanger and E Meyer (Heidelberg: Springer) p 193
[6] Loppacher C, Guggisberg M, Pfeiffer O, Meyer E, Bammerlin M, Lüthi R, Schlittler R, Gimzewski J, Tang H and Joachim C 2003 *Phys. Rev. Lett.* **90** 066107
[7] Bennewitz R, Foster A, Kantorovich L, Bammerlin M, Loppacher C, Schär S, Guggisberg M, Meyer E, Güntherodt H-J and Shluger A 2000 *Phys. Rev. B* **62** 2074

[8] Loppacher C, Bammerlin M, Guggisberg M, Schär S, Bennewitz R, Baratoff A, Meyer E and Güntherodt H-J 2000 *Phys. Rev.* B **62** 16944
[9] Jung T, Schlittler R and Gimzewski J 1997 *Nature* **386** 696
[10] Bethge H and Heydenreich J (ed) 1987 *Electron Microscopy in Solid State Physics* (New York: Elsevier)
[11] Yamamoto K, Iijima T, Kunishi T, Fuwa K and Osaka T 1989 *J. Cryst. Growth* **94** 629
[12] Bennewitz R, Schär S, Barwich V, Pfeiffer O, Meyer E, Krok F, Such B, Kolodzej J and Szymonski M 2001 *Surf. Sci.* **474** L197
[13] Stöhr M, Gabriel M and Möller R 2002 *Europhys. Lett.* **59** 423
[14] Loppacher C, Bennewitz R, Pfeiffer O, Guggisberg M, Bammerlin M, Schär S, Barwich V, Baratoff A and Meyer E 2000 *Phys. Rev.* B **62** 13674
[15] Schär S, Bennewitz R, Eguchi T, Gnecco E, Pfeiffer O, Nony L and Meyer E 2003 *Appl. Surf. Sci.* **210** 43
[16] Forrest S 1997 *Chem. Rev.* **97** 1793
[17] Jung T, Schlittler R, Gimzewski J, Tang H and Joachim C 1996 *Science* **271** 181
[18] Weckesser J, Vita A D, Barth J V, Cai C and Kern K 2001 *Phys. Rev. Lett.* **87** 096101
[19] Eremtchenko M, Schaefer J and Tautz F 2003 *Nature* **425** 602
[20] Israelachvili J 1992 *Intermolecular and Surface Forces* 2nd edn (New York: Academic) chapter 11
[21] Hug H and Baratoff A 2002 *Noncontact Atomic Force Microscopy* (*NanoScience and Technology*) ed S Morita, R Wiesendanger and E Meyer (Berlin: Springer) pp 395–432